# A universal electromagnetic energy conversion adapter based on a metamaterial absorber

**Authors:** Yunsong Xie[1], Xin Fan[1], Jeffrey D. Wilson[2], Rainee N Simons[2], Yunpeng Chen[1], and John Q. Xiao[1]*

**Affiliations:**

[1]Department of Physics and Astronomy, University of Delaware, Newark, Delaware 19716 USA

[2]Glenn Research Center, National Aeronautics and Space Administration, Cleveland, Ohio 44135 USA

*Correspondence to: jqx@udel.edu

**Abstract:** On the heels of metamaterial absorbers (MAs) which produce near perfect electromagnetic (EM) absorption and emission, we propose a universal electromagnetic energy conversion adapter (UEECA) based on MA. By choosing the appropriate energy converting sensors, the UEECA is able to achieve near 100% signal transfer ratio between EM energy and various forms of energy such as thermal, DC electric, or higher harmonic EM energy. The inherited subwavelength dimension and the EM field intensity enhancement can further empower UEECA in many critical applications such as energy harvesting, solar cell, and nonlinear optics. The principle of UEECA is understood with a transmission line model, which further provides a design strategy that can incorporate a variety of energy conversion devices. The concept is experimentally validated at a microwave frequency with a signal transfer ratio of 96% by choosing an RF diode as the energy converting sensor.

**Main Text:**

Electromagnetic (EM) energy conversion is crucial in many applications such as energy harvesting, nonlinear optics, solar cells, etc. A typical realization requires an energy conversion device and an energy conversion adapter. The former converts the incident EM energy into other forms of energy, such as RF diodes in EM energy harvesting applications and silicon PN junctions in solar cell applications; the latter effectively transfers the signal between the incident EM wave and the conversion device, such as antenna structures in energy harvesting applications and antireflection microstructures in solar cell applications. Currently, no energy conversion adapter can be universally adapted to the EM energy conversion devices for realizing versatile purposes in the entire EM spectrum. Antenna structures are widely used in energy harvesting applications to maximize the energy transfer ratio. However, the unavoidable impedance matching circuits [1, 2] required for high signal transfer ratio cause their effectiveness to deteriorate above microwave frequencies. Antireflection microstructures are commonly used in solar cells to optimize the signal transfer ratio by eliminating the reflection wave. However, these structures are not able to enhance the field intensity at selected locations and usually are much thicker at about three times the wavelength of interest [3,4]. Both features limit their usage in miniaturize devices where efficiency depends on the field intensity. Some sub-wavelength structures including nanoantennas have been adapted in nonlinear optical devices to enhance the EM field intensity [5,6]. However, these devices are not able to utilize or confine all incident wave power, which reduces the energy conversion ratio.

One promising candidate that potentially solves all these impediments is a metamaterial absorber (MA). The MA, which typically has subwavelength structure in all dimensions, has been demonstrated to achieve near 100% signal transfer ratio between EM energy and thermal energy [7], to confine EM waves and dramatically enhance the field intensity [8], and to operate over a broad frequency range from microwave [7] to terahertz [9] and optical regions [10-12]. Despite significant progress in this area, most researchers have only focused on using the MA for conversions between electromagnetic and thermal energy [10-14]. A recent report has shown the feasibility of converting the incident EM wave into DC voltage [15]. The detection circuit is rather complicated including a balun, impedance matching circuit, signal amplifier and detection circuit. Consequently, this method may be difficult to be implemented in higher frequency regions such as THz and optical.

In this letter, we report a concept of using an MA as a universal EM energy conversion adapter (UEECA) for converting EM energy into other forms over an entire EM spectrum. An ideal 100% signal transfer ratio can be achieved with an appropriate energy converting sensor. To unravel the underlying principle, we elaborate each MA component by developing a transmission line (TL) model using classical antenna theory with infinite periodic boundary conditions. As a result, a conceptual energy converting sensor, representing the part of incident energy that is usefully converted, can be incorporated into the system. The developed TL model leads to a clear structural design strategy for achieving near 100% signal transfer ratio. The concept of UEECA along with the design strategy is confirmed by an experimental demonstration of converting incident EM wave into DC voltage choosing RF diodes as the energy converting sensor.

The main component of the MA is a periodically distributed metal front structure represented by the yellow dipoles in Fig. 1 (a). Besides the front structure, a separation layer supporting the front structure and a metal back layer is also crucial for constructing a functional MA unit cell as shown in Fig. 1 (b). For a resonant antenna with periodically distributed front structure (Fig. 1 (c)), the components used to model the antenna are the effective capacitance $C_A$, inductance $L_A$, thermal dissipation load $Z_A^T$ and radiation load in both front and back directions $Z_{A\pm}^{Rad}$[16-19]. They are respectively determined by the power stored in the electric field $\widetilde{W_e}$, stored in the magnetic field $\widetilde{W_m}$, dissipated by thermal loss $P^T$ and radiated to infinite space in the front and back direction of the front structure $P_\pm^{Rad}$ (The derivation can be found in the supplementary material). A conceptual electromagnetic energy converting sensor modeled by a load with impedance $Z_L$ is attached on the front structure of every unit cell. This front structure can be characterized using classical antenna theory [19]. According to the model in Fig. 1 (c), the maximum signal transfer ratio between the incident wave and the sensor is limited at $\sqrt{Z_{A>}^{rad}/(Z_{A+}^{rad} + Z_{A-}^{rad})}$ ($Z_{A>}^{rad}$ is the larger value between $Z_{A+}^{rad}$ and $Z_{A-}^{rad}$).

Previous studies suggest that the addition of the metal back plate does not produce serious distortion to the EM field distribution in the front structure, therefore adding the metal back layer will not significantly change the effective RLC components [20]. As a result, the TL model of the MA can be reconstructed as shown in Fig.1 (d). The permittivity $\varepsilon_d$, permeability $\mu_d$, and thickness $h_d$ of the separation layer are effectively described by the components of $TL_S$. In the case where the separation layer is composed of multiple dielectric layers, a group of series-connected $TL_S$ components should be used in the TL model corresponding to each dielectric layer.

According to the model in Fig. 1 (d), the perfect absorption and the maximum transmission between the incident wave and the energy converting sensor can be simultaneously achieved only when:

$$\frac{Z_L^A Z_{TL} + Z_{LC}(Z_{TL} - Z_{A+}^{rad}) - Z_{A+}^{rad}(Z_L^A + Z_{TL})}{Z_L^A Z_{TL} + Z_{LC}(Z_{TL} + Z_{A+}^{rad}) + Z_{A+}^{rad}(Z_L^A + Z_{TL})} = 0 \qquad (1)$$

where $Z_{TL} = i\frac{Z_0 \omega \mu_d h_d}{c}$, $Z_{LC} = i\omega L_A + \frac{1}{i\omega C_A}$ and $Z_L^A = Z_L + Z_A^T$. For physically meaningful solutions, the condition of $Re[Z_L^A] < Z_{A+}^{rad}$ has to be satisfied. By solving Eq. (1), one is able to obtain the smallest separation layer thickness and the corresponding angular frequency, respectively:

$$h_d = \frac{2cL_A C_A \sqrt{Z_L^A}}{\mu_d \sqrt{Z_{A+}^{rad} - Z_L^A} \left[ \sqrt{4L_A C_A + C_A^2 Z_L^A (Z_{A+}^{rad} - Z_L^A)} - \sqrt{C_A^2 Z_L^A (Z_{A+}^{rad} - Z_L^A)} \right]} \qquad (2)$$

$$\omega = \frac{C_A\sqrt{Z_L^A(Z_{A+}^{rad}-Z_L^A)} - \sqrt{4L_A C_A + C_A^2 Z_L^A(Z_{A+}^{rad}-Z_L^A)}}{2C_A L_A} \quad (3)$$

where $c$ is the speed of light in vacuum. These two equations indicate that for the resonant front structure with arbitrary shape, the full absorption can always be achieved by intentionally choosing the thickness of the separation layer. In the meantime, the transmission between the incident wave (emitted wave for the emission process) and the sensor, which is the signal transfer ratio, is optimized to be:

$$T^{Max} = \frac{R_L}{Z_A^T + R_L} \quad (4)$$

For the ideal situation in which the thermal loss can be ignored ($Z_A^T \to 0$), 100% of the signal transfer ratio can be achieved. In practice, $Z_A^T$ is directly related to the dissipation property of the materials as well as the EM field distribution, as indicated by Eq. S6 in the supplementary material. In this case, the largest signal transfer ratio of the device is determined by Eq. (4).

A simple physics picture of the optimum signal transfer ratio can be offered by the interference theory [21], in which the perfect absorption arises from the destructive interference between the direct reflection from the front structure and the multi-reflections inside the separation layer. An appropriate thickness of the separation layer as indicated in Eq. (2) is critical in producing the perfect absorption because the destructive interference is mainly controlled by the phase shift inside the separation layer. What makes UEECA distinct from the conventional MA is how the trapped energy is converted. Instead of being dissipated largely in the separation layer [12] as in a conventional MA design, the existence of the load in a UEECA offers another path from which the transfer ratio efficiency can be determined by Eq. (4).

To validate the concept of a UEECA, an MA array consisting of 12×12 unit cells has been constructed and characterized. An RF diode (Avago HSMS-286B), serving as the energy converting sensor, is attached on each unit cell as shown in Fig. 2(a). The MA is essentially constructed of 4 layers (Fig. 2 (b)): a copper ($5.8 \times 10^7$ S/m in conductivity) front structure (17 μm in thickness) fabricated using standard photolithography techniques; a Rogers 5880LZ first separation layer (0.79 mm in thickness); an air second separation layer (0.80 mm in thickness); and a copper back layer (17 μm in thickness). The real part and loss tangent of the permittivity of the Rogers 5880LZ are 1.96 and 0.002, respectively. The simulated reflected and transferred (to RF diode) signal ratios using HFSS are shown in Fig. 2 (d), where both near zero reflection (-25 dB) and peak transfer (-0.30 dB or 97%) occurs at about 6.7 GHz. The corresponding experimental data are shown in Fig. 2 (e). The minimum reflection of -17 dB and the peak transfer of -0.38 dB or 96% appear at about 6.9 GHz, which agrees well with the simulation results. The measuring method is shown in detail in the supplementary material.

The electric current distribution on the metal structure at the full absorption frequency is shown in Fig. 3 (a), which is similar to the current distribution of a dipole antenna [22]. The largest current flow is located at the center where the sensor is placed. The streamline of the Poynting vector of the incident EM wave is shown in Fig. 3 (b)-(d) in both perspective view and side views. It is clear that most of the incident EM energy is focused, confined and transferred into the lumped port which represents the energy converting sensor.

With a given energy converting sensor, the maximum signal transfer ratio is determined by the thermal loss component $Z_A^T$ in Eq. (4), which comes from both the non-zero permittivity loss tangent of the first separation layer and the finite conductivity of the metal. To illustrate how the optimum signal transfer ratio varies with the loss tangent of the separation layer, a series of simulations using different material as the separation layer have been carried out. A UEECA using FR-4 (with $\varepsilon_r = 4.4\text{-j}0.088$) has also been constructed and tested (see supplementary material). Fig 2 (f) shows the maximum signal transfer ratio as a function of the loss tangent of different materials for both simulation and experimental results (blue). The maximum signal transfer ratio approaches 100% when perfect electric conductor (PEC) is used and the loss tangent of the separation layer material is below $10^{-3}$.

While this experimental demonstration does not include the EM wave polarization and incident angle dependence, the solutions to these problems have already been provided. Absorbing an arbitrary polarization can be accomplished by adopting a 90° rotation symmetrical front structure [21]. For incident angle dependence, the MA maintains near 100% absorption for a TE mode with incident angle between -80° to 80° and above 89% absorption for a TM mode with incident angles smaller than 50° [24]. Therefore, the proposed UEECA is capable of maintaining high signal transfer ratio with different EM wave polarizations and very wide incident angles.

A significant feature of the UEECA is that it can be incorporated with a variety of energy converting sensors. Their different characteristic impedances can be easily accommodated by choosing the appropriate separation layer thickness using Eq. (2). The amount of the incident energy that transfers to the sensor can be maximized and quantitatively calculated using Eq. (4). The schematics of energy conversion process using UEECA can be seen in Fig. 4. The incident EM wave is maximally transferred to different energy converting sensors indicated by $Z_L$ in the TL model of the MA.

The proposed UEECA is feasible for wide applications by accordingly choosing the electromagnetic energy converting device or material as the conceptual sensor $Z_L$. For example, the UEECA eliminates the necessity of an impedance matching circuit for the energy harvesting application. The optimized signal transfer ratio can be easily achieved by choosing the appropriate thickness of the separation layer as suggested by Eq. (3). The UEECA could also make a high impact in solar cell applications. As it has been reported, an MA is able to minimize the thickness of the solar cell to much less than the wavelength and maintain near 100% broadband absorption [28]. The UEECA for solar cell application can be constructed by incorporating the PN junction into the MA as the separation layer. According to Eq. (4), the optimized energy conversion ratio of the solar cell can be achieved as $T^{Max}\beta_{PN}$, where $\beta_{PN}$ is the energy conversion ratio of the PN junction. The UEECA could also be highly desirable in nonlinear optical devices. It advances the previously reported nanostructures [5,6] since the UEECA not only fully absorbs but also confines the incident wave in the region where largest field intensity is located [8]. Dramatic increases of the energy conversion ratio can be realized by integrating a material with nonlinear optical properties as the UEECA separation layer. For instance, $LiNbO_3$ can be selected for second harmonic generation applications.

In summary, by applying the classical antenna theory with infinite periodic boundary conditions, a conceptual energy converting sensor can be integrated into the TL model of a metamaterial absorber (MA). This integration leads to the concept of a universal electromagnetic energy conversion adapter (UEECA). The signal transfer ratio between the incident wave and the sensor can be maximized by simply choosing the optimal separation layer thickness. Using low loss separation material, a UEECA with 96% signal transfer ratio has been experimentally demonstrated.

**ACKNOWLEDGEMENT**

The authors thank the discussions with Dr. Qi Lu. The work is supported from the NASA EPSCoR program under grant number NNX11AQ29A.

**FIGURE CAPTIONS**

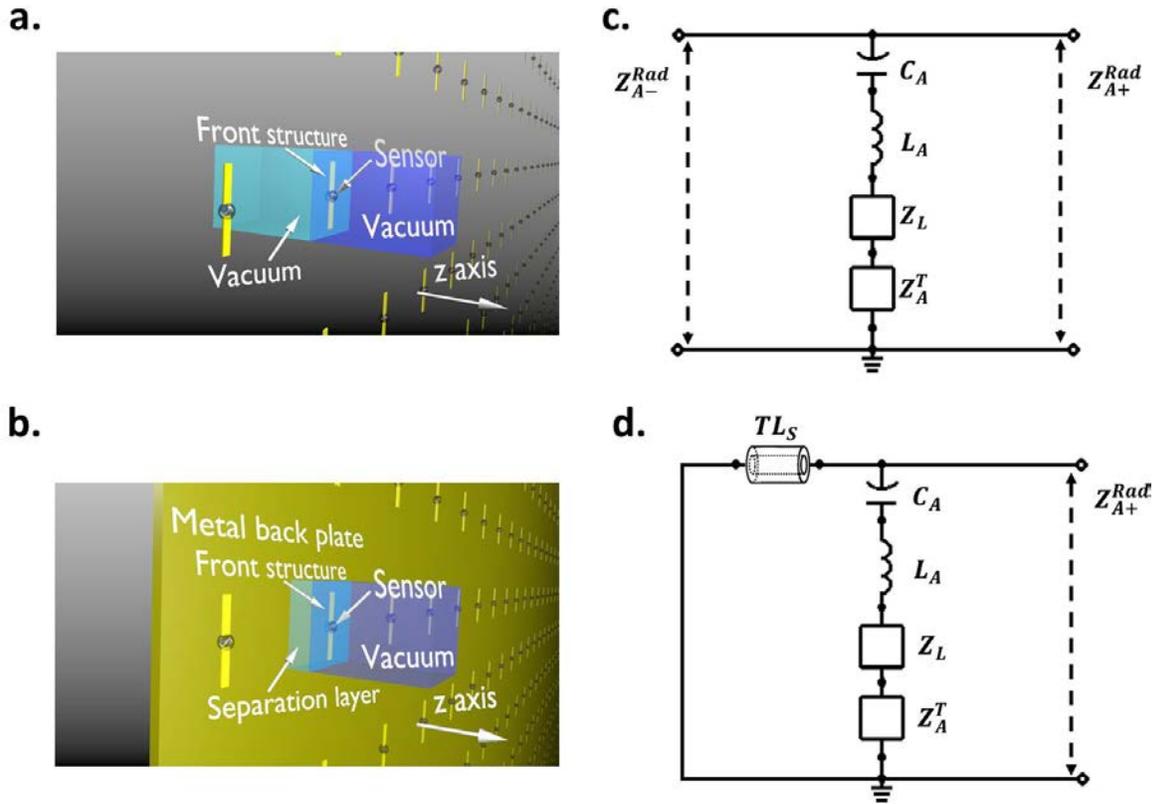

Figure 1. Schematics and TL models for the array of both the front structure and entire MA. The schematics of (a) an array with only front structures and (b) an MA array. The TL model for (c) the front structure of the MA and (d) the TL model for the entire MA which includes the effective capacitance $C_A$, inductance $L_A$, thermal dissipation load $Z_A^T$, and radiation load in both front and back side $Z_{A\pm}^{Rad}$. The shorted line on the far left side represents the metal back plate. The component TLs characterize the shifted phase inside the separation layer.

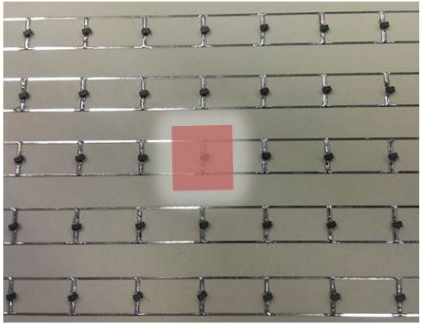
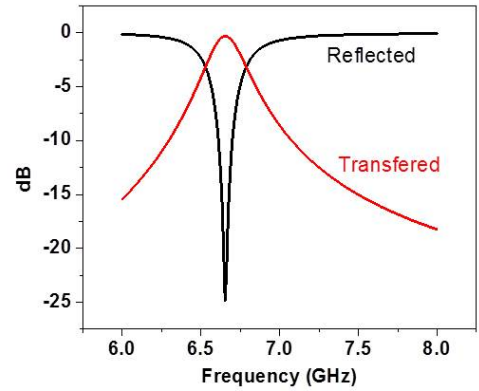
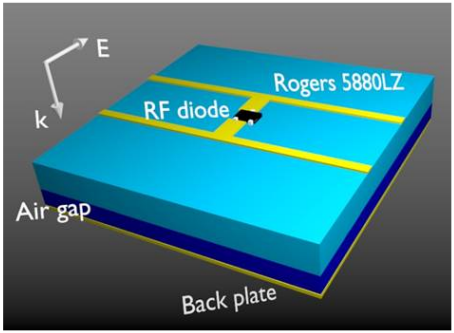
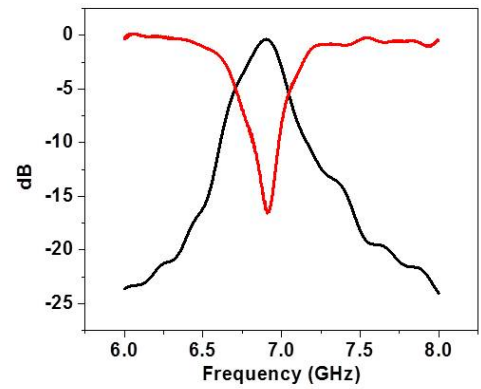
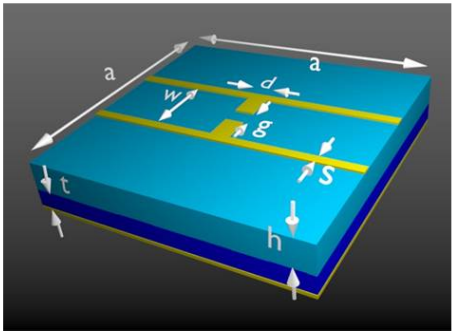
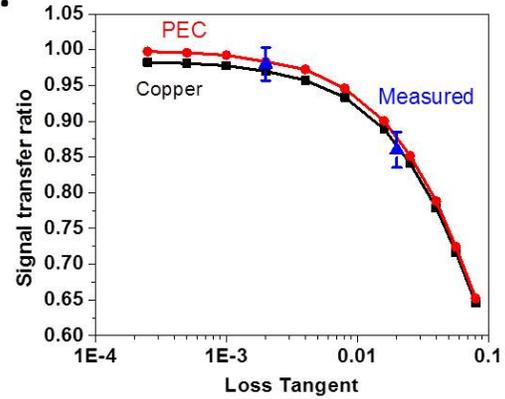

Figure 2. Schematics of the UEECA sample in the measurement, signal transfer ratio spectrum in both simulations and measurement. (a) The front view of the MA sample with RF diodes attached to each unit (highlighted regime). (b) Unit cell components and electric field. (c) Dimensions of a MA unit cell where $a$=15 mm, $d$=1 mm, $g$=2 mm, $s$=0.5 mm and $w$=7 mm.

(d) The HFSS simulated ratio of the reflected signal (black line) and the ratio of the signal being transferred to the RF diode (red line) in the frequency spectrum. (e) The corresponding experimental results. (f) The simulated maximum signal transfer ratio dependence of the permittivity loss tangent of the first separation layer. The material of the metal structure is chosen to be either copper (black line and markers) or PEC (red line and markers). The structural dimensions and the other material parameters are the same as in the previous simulation. The measured maximum signal transfer ratio using FR-4 and Rogers 5880LZ are plotted in (blue up triangles)

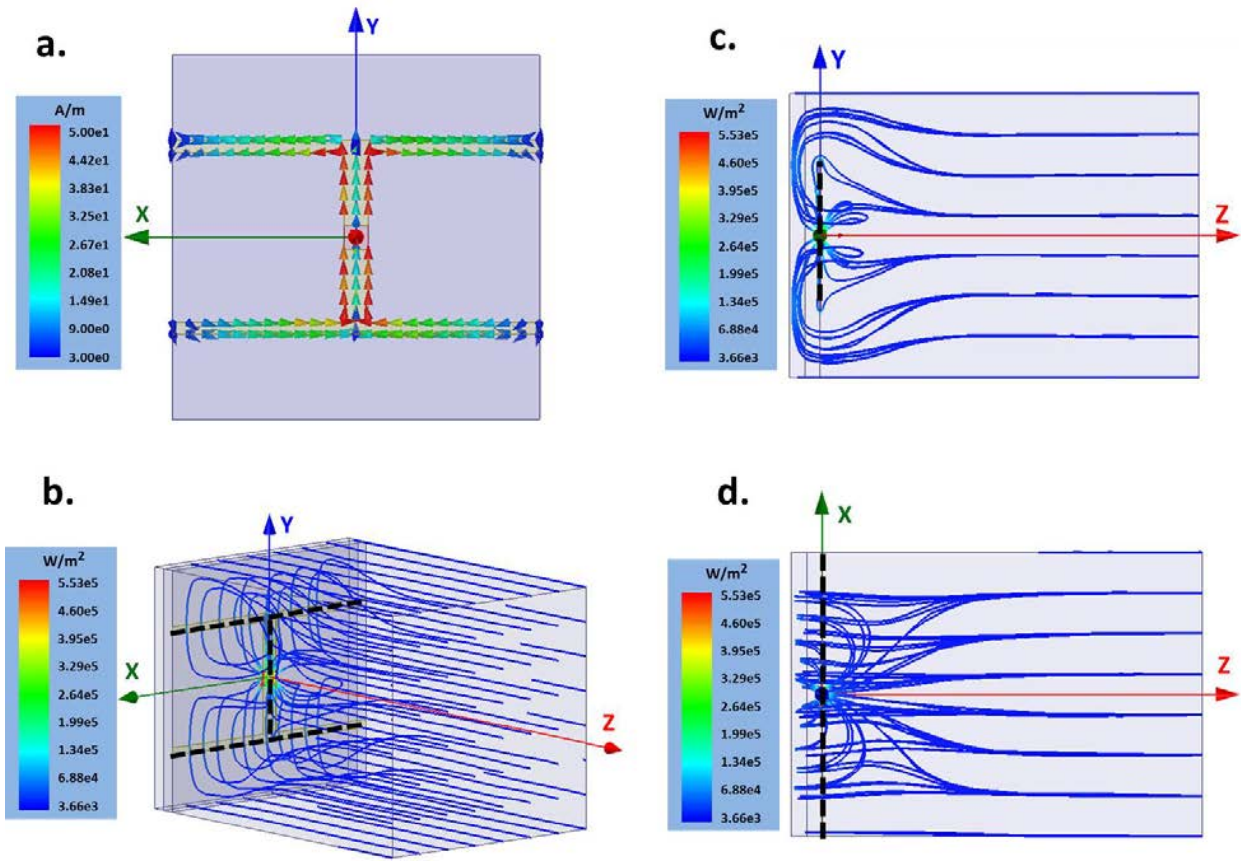

Figure 3. (a) The current distribution on the metal. The (b) perspective view, (c) side view, and (d) top view of the streamlines of the Poynting vector in the UEECA. All the plots are generated

using HFSS at the full absorption frequency. The black dashed lines in (b)-(d) indicate the metal structure; the lumped port at the center represents the energy converting sensor.

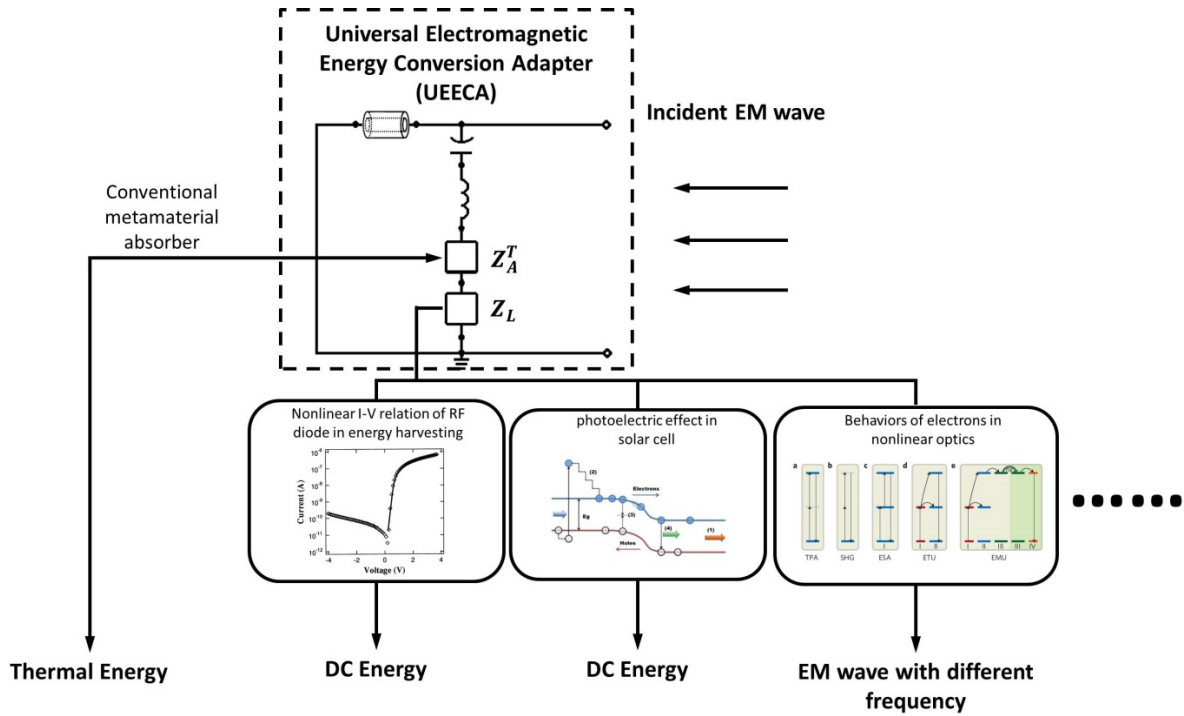

Figure 4. Schematics of the energy flow in the energy conversion processes using the MA-based UEECA. Examples of various energy conversion processes includes the potential subset applications such as energy harvesting [25], solar cells [26], and nonlinear optics [27]

**Measurement setup**

In the measurement, two horn antennas for emission and detection are connected to an Agilent 8722ES network analyzer and are positioned in front of a sample in the far field zone of the antennas. The reflection of the sample is calculated by $R_{Sample} = S_{Sample}/S_{Metal}$, where $S_{Sample}$ and $S_{Metal}$ are the measured transmission between the two antennas when the MA sample and a thick metal plate are, respectively, placed at the same position in front of the antennas. The signal transfer ratio between the RF diode and the incident wave was determined by calculating $\sqrt{P_m/P_i}$, where $P_i$ and $P_m$ are the incident power and the power detected by the RF diode, respectively. The incident power $P_i$ is determined by measuring the transmission between the emission and detection antennas by moving the detection antenna to the sample location. The RF diode detected power $P_m$ is calculated by $V_m/\beta$, where $V_m$ is the output DC voltage from the parallel connected RF diode and $\beta$ is the signal sensitivity of the RF diode. The DC voltage from the RF diodes is measured by a Keithley 2000 multimeter.

The simulation of the UEECA is carried out by the commercial HFSS simulation software. Master and slave boundaries are applied on faces parallel to the EM wave propagation direction to mimic the periodically distribution of the units. Floquet excitation is used to simulate the incident wave. More than 20000 tetrahedrons in the final mesh process are used in the model.

**Measurement result of the UEECA constructed using FR4**

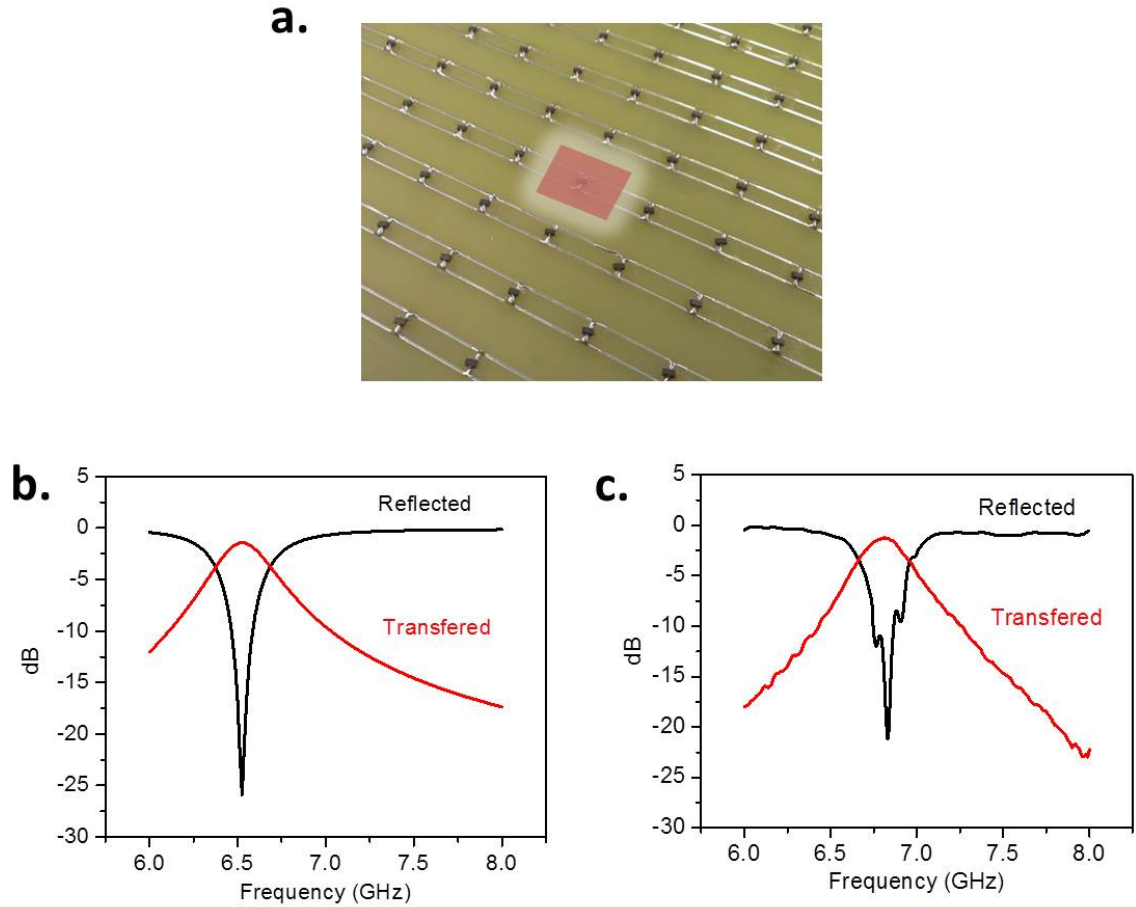

Figure.S1 (a) The picture of the UEECA constructed using FR4, the ratio of the reflected and transferred microwave in (b) simulation and (c) measurement.

FR-4 has been used to construct a demonstration of UEECA as shown in Fig. S1 (a). This array is formed with 12 by 12 units. Avago HSMS-286B diode is attached on every unit of the UEECA. The dimensional parameters of the unit are $a$=15 mm, $w$=4 mm, $s$=0.5 mm, $d$=1 mm, $g$=2 mm, $h$=1.58 mm and $t$=1 mm. The real part and loss tangent of the permittivity of Rogers 5880LZ are set to be 4.4 and 0.02 respectively. A maximum signal transfer ratio of 87% (-1.2 dB) has been achieved in a simulation at frequency 6.5 GHz, where the reflection is -26 dB. The experimental measurement of the sample gives maximum signal transfer ratio of 86% (-1.3 dB) at 6.8 GHz, where the reflection reads -23 dB.

**Derivation of the transmission line model of the front structure of MA in the periodic boundary condition**

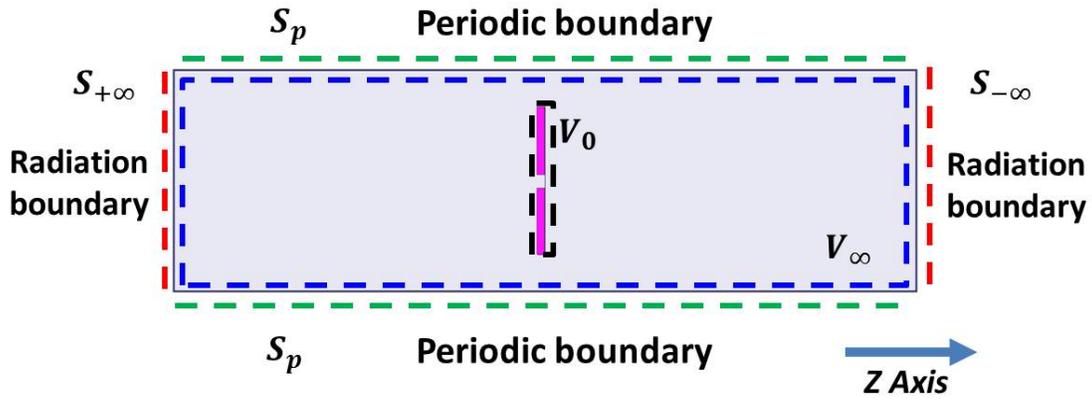

Figure.S2 A sketch of the structure with only the front structure as in Fig. 1 (a). The notations include the radiation boundary surfaces in the positive (negative) z axis $S_{+\infty}(S_{-\infty})$ indicated by the red dashed lines, periodic boundary surfaces $S_p$ shown by the green dashed lines, far field volume $V_\infty$ surrounded by the blue dashed line, and the volume containing only the antenna $V_0$ surrounded by the black dashed line.

The periodically distributed arbitrary front structure is considered as a resonant antenna as shown in Fig. S2. A conceptual energy converting sensor modeled by a load with impedance $Z_L$ is attached to the front structure of every signal unit. The value of $Z_L$ is:

$$Z_L = \frac{2P_L}{|I|^2} \qquad (S1)$$

where $P_L$ is the converted EM power or the power being emitted from the sensor and $I$ is the intensity of the current exciting the antenna. The rest of the components in Fig. 1 (c) can be expressed by[1]:

$$Z_A^T = \frac{2P^T}{|I|^2} \qquad (S2)$$

$$L_A = \frac{4\widetilde{W}_m}{|I|^2} \qquad (S3)$$

$$C_A = \frac{|I|^2}{4\omega^2 \widetilde{W}_e} \qquad (S4)$$

The radiation loads in the positive (+) and negative (-) z axis directions can be expressed as:

$$Z_{A\pm}^{Rad} = \frac{2}{|I|^2} \frac{\left(P_{A+}^{Rad} + P_{A-}^{Rad}\right)^2}{P_{A\pm}^{Rad}} \qquad (S5)$$

where $Z_A^T$ and $Z_{A\pm}^{Rad}$ are the impedances representing the thermal dissipation and radiation, and $L_A$ and $C_A$ are the inductive and capacitive components of the antenna. $P^T$, $P_{\pm}^{Rad}$, $\widetilde{W}_m$ and $\widetilde{W}_e$ are defined under the condition that the antenna is excited by a current $I$. $P^T$ represents the power dissipated and converted into thermal energy because of the non-zero imaginary part of either permittivity or permeability of the substrate and ohmic losses in the non-perfect metal conductor. $P_{\pm}^{Rad}$ is the power being radiated to the left (-) and right (+) sides of the structure, which are the same for an antenna of symmetric structure as shown in Fig. 1 (a) and Fig. S2. $\widetilde{W}_m$ and $\widetilde{W}_e$ stand for the stored magnetic and electric field power, respectively. The values of $P^T$, $P_{\pm}^{Rad}$, $\widetilde{W}_m$ and $\widetilde{W}_e$ can be calculated by[2,3]:

$$P^T = \frac{\omega}{2} \int_{V_\infty} \varepsilon''|\boldsymbol{E}|^2 + \mu''|\boldsymbol{H}|^2 \, dv \quad (S6)$$

$$P_\pm^{Rad} = Re \oint_{S_{\pm\infty}} (\boldsymbol{E} \times \boldsymbol{H}^*) \cdot d\boldsymbol{S} \quad (S7)$$

$$\widetilde{W_m} =$$

$$\frac{1}{2}\left[\int_{V_\infty - V_0} \frac{1}{4}(\varepsilon'|\boldsymbol{E}|^2 + \mu'|\boldsymbol{H}|^2) dv - \frac{r_\infty}{c} Re \oint_{S_{-\infty} + S_{+\infty} + S_p} (\boldsymbol{E} \times \boldsymbol{H}^*) \cdot d\boldsymbol{S} + \frac{1}{2\omega} Im \oint_{S_{-\infty} + S_{+\infty} + S_p} (\boldsymbol{E} \times \boldsymbol{H}^*) \cdot d\boldsymbol{S}\right] \quad (S8)$$

$$\widetilde{W_e} =$$

$$\frac{1}{2}\left[\int_{V_\infty - V_0} \frac{1}{4}(\varepsilon'|\boldsymbol{E}|^2 + \mu'|\boldsymbol{H}|^2) dv - \frac{r_\infty}{c} Re \oint_{S_{-\infty} + S_{+\infty} + S_p} (\boldsymbol{E} \times \boldsymbol{H}^*) \cdot d\boldsymbol{S} - \frac{1}{2\omega} Im \oint_{S_{-\infty} + S_{+\infty} + S_p} (\boldsymbol{E} \times \boldsymbol{H}^*) \cdot d\boldsymbol{S}\right] \quad (S9)$$

(in my version, integral signs are missing upper limits) where $\boldsymbol{E}$ and $\boldsymbol{H}$ are the electric field and magnetic field distributions in a single unit under the periodic boundary conditions, $\omega$ is the angular frequency of the EM wave, and $\varepsilon'$ ($\mu'$) and $\varepsilon''$($\mu''$) are the real and imaginary part of the permittivity (permeability) of the constituent material, respectively. $V_0$ is the volume that contains the antenna, $V_\infty$ is the volume that covers the region, the edges of which can be viewed as in the far field region, $c$ is the speed of light in vacuum, and $r_\infty$ is the distance between the antenna and the edge of the $V_\infty$ regime. It is reasonable to make the conclusion that there is no net energy flow through the periodic boundary surfaces:

$$\oint_{S_p} (\boldsymbol{E} \times \boldsymbol{H}^*) \cdot d\boldsymbol{S} = \boldsymbol{0} \quad (S10)$$

(upper limit missing) It is worth noticing that the equation above does not indicate that there is no coupling between the units. Instead the Poynting vectors in the perpendicular direction at the periodic boundaries cancel out, which results in a vanishing integral. The calculations of $\widetilde{W_m}$ and $\widetilde{W_e}$ should consider the electric and magnetic power that is stored in the volume of the antenna structure. Therefore, the equations (8-9) can be expressed by:

$$\widetilde{W_m} = \frac{1}{2}\left[\int_{V_\infty} \frac{1}{4}(\varepsilon'|E|^2 + \mu'|H|^2)dv - \frac{r_\infty}{c} Re \oint_{S_{-\infty}+S_{+\infty}} (E \times H^*) \cdot dS + \frac{1}{2\omega} Im \oint_{S_{-\infty}+S_{+\infty}} (E \times H^*) \cdot dS\right] \quad (S11)$$

$$\widetilde{W_e} = \frac{1}{2}\left[\int_{V_\infty} \frac{1}{4}(\varepsilon'|E|^2 + \mu'|H|^2)dv - \frac{r_\infty}{c} Re \oint_{S_{-\infty}+S_{+\infty}} (E \times H^*) \cdot dS - \frac{1}{2\omega} Im \oint_{S_{-\infty}+S_{+\infty}} (E \times H^*) \cdot dS\right] \quad (S12)$$